\documentclass[pra,twocolumn,graphicx,amssymb,floatfix]{revtex4-1}
\usepackage{color}
\usepackage{amsmath}
\usepackage{graphicx}
\usepackage{float}

\definecolor{orange}{rgb}{1,0.5,0}
\definecolor{darkgreen}{rgb}{0.0, 0.578125, 0.25}

\begin{document}

\title{  Weak value beyond conditional expectation value of the pointer readings}
%\centerline{}
%\centerline{}
\author{Lev Vaidman$^1$, Alon Ben-Israel$^1$, Jan Dziewior$^{2,3}$, Lukas Knips$^{2,3}$,  Mira Wei{\ss}l$^{2,3}$, Jasmin Meinecke$^{2,3}$, Christian Schwemmer$^{2,3}$, Ran Ber$^1$, and Harald Weinfurter$^{2,3}$}
\affiliation{ $^1$ Raymond and Beverly Sackler School of Physics and Astronomy,
 Tel-Aviv University, Tel-Aviv 69978, Israel\\$^2$ Max-Planck-Institut f\"{u}r Quantenoptik, Hans-Kopfermann-Stra{\ss}e 1, 85748 Garching, Germany\\
$^3$ Department f\"{u}r Physik, Ludwig-Maximilians-Universit\"{a}t, 80797 M\"{u}nchen, Germany\\
}

\begin{abstract}
It is argued that a weak value of an observable is a robust property of a single pre- and post-selected quantum system rather than a statistical property.
During an infinitesimal time a system with a given weak value affects other systems as if it were in an eigenstate with eigenvalue equal to the weak value.
This differs significantly from the action of a system pre-selected only and possessing a numerically equal expectation value.
The weak value has a physical meaning beyond a conditional average of a pointer in the weak measurement procedure.
The difference between the weak value and the expectation value has been demonstrated on the example of photon polarization.
In addition,  the  weak values for systems pre- and post-selected in mixed states are considered.
\end{abstract}
\maketitle

\section{Introduction} \label{introduction}

Contrary to classical physics, in quantum mechanics physical observables or, in short, variables might not have definite values even if the full quantum description of the system  is given.
If the system is in a superposition or a mixture of eigenstates corresponding to different eigenvalues, then only an expectation value of a variable is defined.
But since, in general, this value cannot be measured given a single system, the expectation value is commonly associated  with an ensemble of identically prepared systems and thus not considered as a property of a single system.
If, however, the system is in an eigenstate of the observable, a measurement on a single system is enough to observe the corresponding eigenvalue.
It is therefore possible to attribute the eigenvalue to a single quantum system.

Again, contrary to classical physics, even if we have complete information about the preparation of a system, the outcomes of future  measurements are in general not determined.
The two-state vector formalism \cite{ABL,AVreview} uses such future results to describe systems in between two measurements. It employs two quantum states, the usual one defined by the preparation and the backward evolving state defined by the result of the post-selection.
For describing the coupling to variables of a pre- and post-selected system the concept of a {\it weak value} has been introduced  \cite{AAV}.
It was accepted with harsh criticism \cite{AAVLeg,AAVPer,AAVrep}, but its significance was appreciated in a number of experiments performing weak measurements in a regime of anomalous weak values achieving unprecedented precision in measuring small parameters \cite{Kwiat,Dixon}.
New proposals and experiments continue to appear \cite{Stru,Xu,Zhou,JaMe,Boyd,cycle}.
Still, a controversy regarding the usefulness of anomalous weak values for parameter estimation and metrology compared to conventional methods based on strong measurements arose \cite{FC1,Knee,JoHo,Das,Lee,PaBrPRL,ZDW,Nish,Al-Da,PaBrPRA,SuTa,ViHo}.
Moreover, even the quantum nature of weak values was questioned \cite{FC2,Brod,Pu,Dr-No,Karanj,Dress,Soc,Ip}.

Weak values were introduced as outcomes of weak measurements \cite{AAV}, which have large uncertainty in the pointer position.
Thus, in experiments, the weak value is obtained as a statistical average of the pointer readings.
Even among proponents of this concept the weak value is frequently understood as a  mere generalization of the expectation value for the case when the quantum system is post-selected, i.e., a conditional expectation value \cite{review}.

In this paper we argue that although we obtain weak values from many measurement outcomes, they represent definite properties of single pre- and post-selected quantum systems.
Furthermore, we show that a weak value is more similar to an eigenvalue than to an expectation value.

This holds when the system is described by a pure two-state vector and also when it is described by a {\it generalized two-state vector} \cite{AV91}.
In addition, we generalize the definition of weak values for genuine mixed pre- and post-selected states \cite{SANDU}.
Evidently, then, but only then, the weak value has the nature of a statistical average.

\section{Comparing weak values and expectation values} \label{first_example}

The weak value of a variable $A$ is defined for a single quantum system pre-selected in a state $ |\psi \rangle$ and post-selected in a state $ |\varphi \rangle$ as
\begin{equation}\label{wv}
A_w\equiv\frac{\langle \varphi |A|\psi\rangle}{\langle \varphi |\psi\rangle}.
\end{equation}

In our analysis we stay within the framework of the standard quantum theory describing the system and systems it interacts with using wave functions only.
We view the weak value of a variable of a system as a description of its interaction with other systems, i.e., the way the pre- and post-selected system influences the quantum state of other systems.

While the definition of the weak value is based on the pure two-state vector with states $ |\psi \rangle$ and $ |\varphi \rangle$  considered at a particular time $t$, its operational meaning relies on interactions with other systems which create entanglement.
The way to deal with this problem is by a delicate play of limits.
We should consider a short period of time around time $t$ to evaluate the action of the system on other systems.
The change in the other systems should be large enough to be seen, but the back action on the system should be small enough, such that the change in the two-state vector describing the system can be neglected.
Since we allow an unlimited ensemble of experiments with identical pre- and post-selection, the required limits are achievable.

We start our analysis with a von Neumann measurement Hamiltonian coupling a quantum system to a continuous pointer system defined by its position $Q$,
\begin{equation}\label{VNInt}
H_{\rm int}= gA P,
\end{equation}
where $P$ is a conjugate momentum to $Q$ and $g$ is a coupling constant. % \COMMENT{with the dimension $[g]=\frac{J}{[P]}$}.
We assume that at time $t=0$, the system was prepared in state $|\psi \rangle$ and shortly after, at time $t=\epsilon$ was found in state $|\varphi \rangle$.
For simplicity of calculations, we assume that the pointer at time $t=0$ is in a Gaussian state
\begin{equation}\label{Phi0}
\Phi_0=\frac{1}{(2\pi)^{1/4}\sqrt{\Delta}}e^{-\frac{Q^{2}}{4\Delta^{2}}}.
\end{equation}

For a comparison of different cases, we consider the pointer state after the interaction with the system and the post-selection measurement at time $t=\epsilon$.
Let us consider a particle with the integer spin observable
\begin{equation}\label{A=SZ}
A\equiv S_z = \sum_j j | j \rangle \langle j |.
\end{equation}
If, e.g., the spin state is the eigenstate $|1\rangle$, i.e. the variable has the eigenvalue $A=1$, then at time $t=\epsilon$, independently of the result of the post-selection measurement, the pointer state is shifted:
\begin{equation}\label{PhiF}
\Phi_{\rm e}=\frac{1}{(2\pi)^{1/4}\sqrt{\Delta}}e^{-\frac{\left(Q-g\epsilon \right)^{2}}{4\Delta^{2}}}.
\end{equation}
where we have set $\hbar \equiv 1$ here and throughout the rest of the paper.
If the system is not in an eigenstate of $A$, the pointer state might also be distorted.

To compare various cases we evaluate the effect of the interaction by calculating the distance between quantum states  expressed by the Bures angle.
The distance between the initial state of the measuring device (\ref{Phi0}) and the final state (\ref{PhiF}) is
\begin{equation} \label{Bures_weak}
D_A\left(\Phi_0,\Phi_{\rm e}\right)\equiv\arccos \left|\left\langle \Phi_0|\Phi_{\rm e}\right\rangle \right|=\frac{g\epsilon}{2\Delta}+\mathcal{O}\left(\epsilon^3\right).
\end{equation}
The amplitude of  the additional orthogonal component is of the order of $\frac{g\epsilon}{2\Delta}$, but since our theoretical small parameter is $~~\epsilon \ll \frac{\Delta}{g}$, we take into account only the lowest order of $\epsilon$.

Consider now a pre- and post-selected system with $A_w=1$, but in which both pre-selection and post-selection do not include the eigenstate $|1\rangle$.
A two-state vector which provides this weak value is
\begin{equation}\label{TSV}
\langle\varphi|~~|\psi\rangle=\frac{1}{\sqrt 5}\left(\langle-1| - 2 \langle0|\right)~~~\frac{1}{\sqrt 2}\left(|-1\rangle+ |0\rangle \right).
\end{equation}
After the post-selection, the state of the pointer variable is
\begin{equation}
\Phi_{\rm w}=\mathcal{N}(\epsilon)(2e^{-\frac{Q^{2}}{4\Delta^{2}}}-e^{-\frac{(Q+g\epsilon)^{2}}{4\Delta^{2}}}) \approx \mathcal{N}^\prime(\epsilon) e^{\frac{ -Q^2g^2\epsilon^2}{4 \Delta^4}} \Phi_{\rm e}.\label{wvpointer}
\end{equation}
Note that because $\Delta^2/g\epsilon\gg\Delta$, the distortion factor $e^{\frac{ -Q^2g^2\epsilon^2}{4 \Delta^4}}$ is almost constant relative to $\Phi_{\rm e}$.
Therefore $\Phi_{\rm w}$ is effectively a Gaussian centered around $A_w=1$ and is thus very close to $\Phi_{\rm e}$ as seen from the Bures angle
\begin{equation} \label{Phi2-Phi1}
D_A\left(\Phi_{\rm e},\Phi_{\rm w}\right)=\frac{g^2\epsilon^2}{2\sqrt{2}\Delta^2}+\mathcal{O}\left(\epsilon^4\right).
\end{equation}

The characteristic distance between states after the interaction for the time $\epsilon$ is approximately $\frac{g\epsilon}{2\Delta}$, so when the additional distance is proportional to $\epsilon^2$, it can be neglected.
Thus, in the limit of short interaction times, the pre- and post-selected system with some weak value interacts with other systems in the same manner as a system pre-selected in an eigenstate with a numerically equal eigenvalue.
Not only the expectation values of the positions of the pointers are essentially the same, but the full quantum states of the pointers are almost identical.

The situation changes considerably when the system is only pre-selected in a state with the expectation value $\langle A\rangle=1$, which, however, is not the eigenstate $|1\rangle$.
To show this, assume that the particle is in the state
\begin{equation}\label{psi3}
|\psi\rangle= \frac{1}{\sqrt 2}\left (| 0\rangle +| 2\rangle \right ).
\end{equation}
At time $t=\epsilon$, now without post-selection, the pointer system is not described by a pure state, but by a mixture.
The density matrix describing this mixture is
\begin{equation}
\rho_{\rm ex} \!=\!\frac{1}{2\sqrt{2\pi}\Delta}\left(e^{-\frac{Q^2+Q'^2}{4\Delta^2}}\!+\!e^{-\frac{(Q-2g\epsilon)^2 +(Q'-2g\epsilon)^2}{4\Delta^2}}\right)\!.\label{mixed}
\end{equation}

The distance between $\rho_{\rm ex}$ and $\Phi_{\rm e}$, the state of the pointer after coupling to an eigenstate (\ref{PhiF}), is
\begin{equation} \label{overlap_mixture}
D_A\left(\Phi_{\rm e},\rho_{\rm ex}\right) \equiv \arccos(\sqrt{\langle \Phi_{\rm e} |\rho_{\rm ex} | \Phi_{\rm e} \rangle})=\frac{g\epsilon}{2\Delta}+\mathcal{O}\left(\epsilon^3 \right).
 \end{equation}

This is a significantly larger distance than (\ref{Phi2-Phi1}).
In fact, the distance (\ref{overlap_mixture}) is of the same order as (\ref{Bures_weak}) and cannot be neglected for small $\epsilon$.

While the pointer states (\ref{wvpointer}) and (\ref{mixed}) correspond to similar probability distributions, they are fundamentally different.
As in the case of an eigenstate (\ref{PhiF}), the final pointer state (\ref{wvpointer}) corresponds to a shift of the original distribution given by a single number, the weak value.
In this sense, we call eigenvalues and weak values ``robust''.
The situation changes significantly when the system is prepared in a superposition of eigenstates.
This results in an incoherent mixture of pointer distributions (\ref{mixed}), which cannot be described by a single parameter anymore, but by several parameters, namely all eigenvalues corresponding to the superposed eigenstates.
Consequently, this pointer is in a statistical mixture of independent distributions in stark contrast to the other cases.
In the above example, the statistical character of the expectation value is reflected in (\ref{mixed}), which represents the mixture of two independent distributions centered around the eigenvalues $0$ and $2$.

Note that if we add post-selection on the original state (\ref{psi3}), such that $A_w=1$, then the coupling of the system to the external system converges to that of the system described by eigenvalue $A=1$, again.

The system evolves as in the eigenvalue case also when the weak value is anomalously large and lies very far from the range of the actual eigenvalues. It might be equal to an eigenvalue which is not present in the pre- and post-selected states, or it might be equal to a non-existing, ``hypothetical'' eigenvalue of the system, as in the canonical example \cite{AAV}.
For example, a system described by the two-state vector
\begin{equation}\label{100}
\langle\varphi|~~|\psi\rangle=\frac{1}{\sqrt{20201}}\left(100\langle-1| - 101 \langle0|\right)~~~\frac{1}{\sqrt 2}\left(|-1\rangle+ |0\rangle \right),
\end{equation}
has a weak value of $100$.
Then, the distance between the states of the external system at time $t=\epsilon$ in this case and in the case when the system has an eigenvalue $A=100$,
\begin{equation}\label{DA_100}
D_A\left(\Phi_{\rm e},\Phi_{\rm w}\right)=\frac{100\cdot101g^2\epsilon^2}{4\sqrt{2}\Delta^2}+\mathcal{O}\left(\epsilon^4\right),
\end{equation}
is larger than (\ref{Phi2-Phi1}), but still scales favorably with $\epsilon^2$.
As will be shown in Sec.~\ref{General}, the resulting scaling order is not restricted to these examples, but holds in general.

\section{Experimental demonstration}

Our claim that the system described by some weak value affects other systems similarly to the
action of a system described by a numerically equal eigenvalue, but differently than a system described by a numerically equal expectation value, is tested experimentally.
 Our system is the polarization degree of freedom of the Gaussian light beam. The ``other'' or ``pointer'' system is the transverse position of the beam.
The variable  $A$ of the system includes the polarization operator ${\mathcal P}$,
\begin{equation}\label{polar}
{\mathcal P}|H\rangle =|H\rangle, ~~{\mathcal P}|V\rangle =-|V\rangle,
\end{equation}
The coupling to this variable is achieved by passing the beam through a birefringent crystal which shifts the beam according to its polarization.
We consider situations in which the weak value and expectation value equal $0$.
The operator ${\mathcal P}$ has only eigenvalues $\pm1$, but effectively we can perform the experiment also for $A$ with the eigenvalue $0$ by removing the birefringent element from the beam or tilting it so that there is no shift.

We start by preparing light in the state
\begin{equation}\label{pre}
|\psi\rangle= \frac{1}{\sqrt 2}\left (| H\rangle +| V\rangle \right )\equiv \frac{1}{\sqrt 2}\left (| 1\rangle +|- 1\rangle \right ).
\end{equation}
Then, effectively, we have the three cases of interest: (i) without the crystal, the \textit{eigenvalue} $A=0$; (ii) with the crystal, pre-selection only of the state (\ref{pre}) with \textit{expectation value} $\langle A\rangle=0$; and (iii) pre- and post-selection of the same state (\ref{pre}) with the \textit{weak value} $A_w=0$.

Coupling to a state with ``eigenvalue'' $0$ causes no change of the state of the pointer system, meaning that the Gaussian beam is not shifted, and  $\Phi_{\rm e}$ has the form of (\ref{Phi0}).

Coupling to the state (\ref{pre}) with expectation value $0$ leads to the mixed state of the pointer described by density matrix
\begin{equation}
\rho_{\rm ex} \!=\!\frac{1}{2\sqrt{2\pi}\Delta}\left(e^{-\frac{(Q+g\epsilon)^2 +(Q'+g\epsilon)^2}{4\Delta^2}}\!+\!e^{-\frac{(Q-g\epsilon)^2 +(Q'-g\epsilon)^2}{4\Delta^2}}\right)\!.\label{mixed1}
\end{equation}

For post-selected systems, the coupling to a state with weak value $0$ leads to the pointer state
\begin{equation}\label{phiw}
\Phi_{\rm w} =\mathcal{N}(e^{-\frac{(Q-g\epsilon)^{2}}{4\Delta^{2}}} + e^{-\frac{(Q+g\epsilon)^{2}}{4\Delta^{2}}}).
\end{equation}

The distances between the states and the reference $\Phi_{\rm e}$ given by (\ref{Phi0}) behave according to the results of the first example (\ref{overlap_mixture}), (\ref{Phi2-Phi1}) and are given by
\begin{eqnarray}
D_{A}\left(\Phi_{{\rm e}},\rho_{{\rm ex}}\right) &	\simeq &	\frac{g\epsilon}{2\Delta}, \label{exp_angle_ex}\\
D_{A}\left(\Phi_{{\rm e}},\Phi_{{\rm w}}\right)	& \simeq &	 \frac{g^{2}\epsilon^{2}}{4\sqrt{2}\Delta^{2}} \label{exp_angle_weak}.
\end{eqnarray}

The center of the beam is the same in all cases, the difference between the states is only due to different distortions of the initial Gaussian beam.
Our method to measure this difference is to perform an interference experiment between the reference beam and the beam distorted by the measurement interaction for the two other cases. It is based on the following simple relation between maximal obtainable visibility of the interference between two beams and the Bures angle between their quantum states
\begin{equation}\label{VisBur}
\arccos{{\cal V}_{\rm max}}= D_{A}.
\end{equation}
This formula holds when the visibility is maximal under variation of all parameters: alignment, relative intensity and - in the case of interference between pure and mixed state - unitary transformation of auxiliary degrees of freedom, the polarization in our case. It follows from the well known expression for the pure states, ${{\cal V}_{\rm max}}=|\langle \Psi|\Phi\rangle|$ and from Uhlmann's theorem \cite{Uhlmann76} about purification of mixed states.

\begin{figure}[h]
\centering
\includegraphics[width=0.48\textwidth]{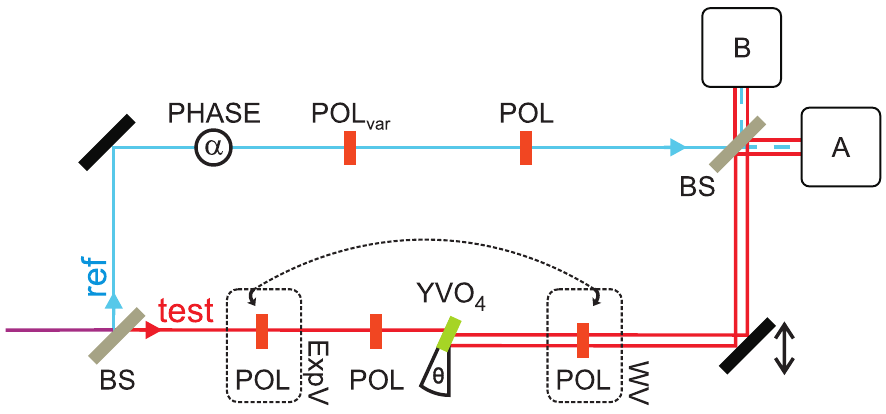}
\caption{Experimental setup for comparing pointer wave functions.
By varying the phase $\alpha$, we can change between constructive and destructive interference in order to measure the maximal visibility and thus to find the Bures angle.
The first polarizer in the reference beam (ref) is used to vary the intensity of the reference beam.
The post-selection polarizer in the test beam (test) used when analyzing the weak value (WV) case is removed and put in front of the YVO$_4$ when observing the expectation value in the test arm (ExpV) to keep the number of optical components equal.
Note that for practical reasons a folded Mach-Zehnder interferometer, i.e., a non-degenerate version of a Michelson interferometer, is used, where the phase can be scanned by means of a retroreflecting prism coupled to a piezo.
The shift of the center of the test beam caused by the YVO$_4$ is compensated by a lateral displacement of the beam.
}
\label{setup}
\end{figure}

 In our experiment we use a balanced Mach-Zehnder interferometer where the incoming beam is preselected in the specified polarization (\ref{pre}), see Fig. \ref{setup}.
In one arm of the interferometer, the spatial mode remains Gaussian ($\lambda=780\,\mathrm{nm}$, beam waist $w=2\Delta=813\,\mathrm{\mu m}$) and can therefore be described by (\ref{Phi0}).
The state in this arm, $\Phi_{\rm e}$, is used as reference.
In the other arm we insert a Yttrium Orthovanadate (YVO$_4$) crystal (of thickness ${d=4.52\,\mathrm{mm}}$), which causes a transversal shift according to the action of ${\mathcal P}$ from (\ref{polar}), thus coupling the (Gaussian) spatial mode of the beam to the different polarizations.
For a tilting angle $\theta$, the beam separation is
\begin{equation}
\delta x  =  d\left(\frac{\sin\left(\theta - \theta^{(o)}\right)}{\cos\left(\theta^{(o)}\right)} - \frac{\sin\left(\theta - \theta^{(e)}\right)}{\cos\left(\theta^{(e)}\right)}\right),
\end{equation}
where $\theta^{(i)} = \arcsin\left(\frac{\sin(\theta)}{n^{(i)}}\right)$ is obtained via Snell's law for $i\in\{o,e\}$ with $o$ ($e$) denoting the ordinary (extraordinary) beam.
Varying the tilting angle $\theta$ mimics varying the interaction time $\epsilon$ according to $\delta x = 2g\epsilon$.

For the expectation value case the system is described by density matrix (\ref{mixed1}).
For the weak value case (\ref{phiw}) we add, behind the YVO$_4$ crystal, a polarization filter aligned in the original polarization state (\ref{pre}).
In order to find the maximal visibility for different interaction strengths, the alignment of the interferometer was optimized and the relative intensities of the arms were varied by means of the first polarizer in the reference arm.
There was no need to vary the reference polarization since in our setup the optimal polarization does not depend on the interaction strength.

\begin{figure}[h]
\label{Exp1}
\centering
\includegraphics[width=0.48\textwidth]{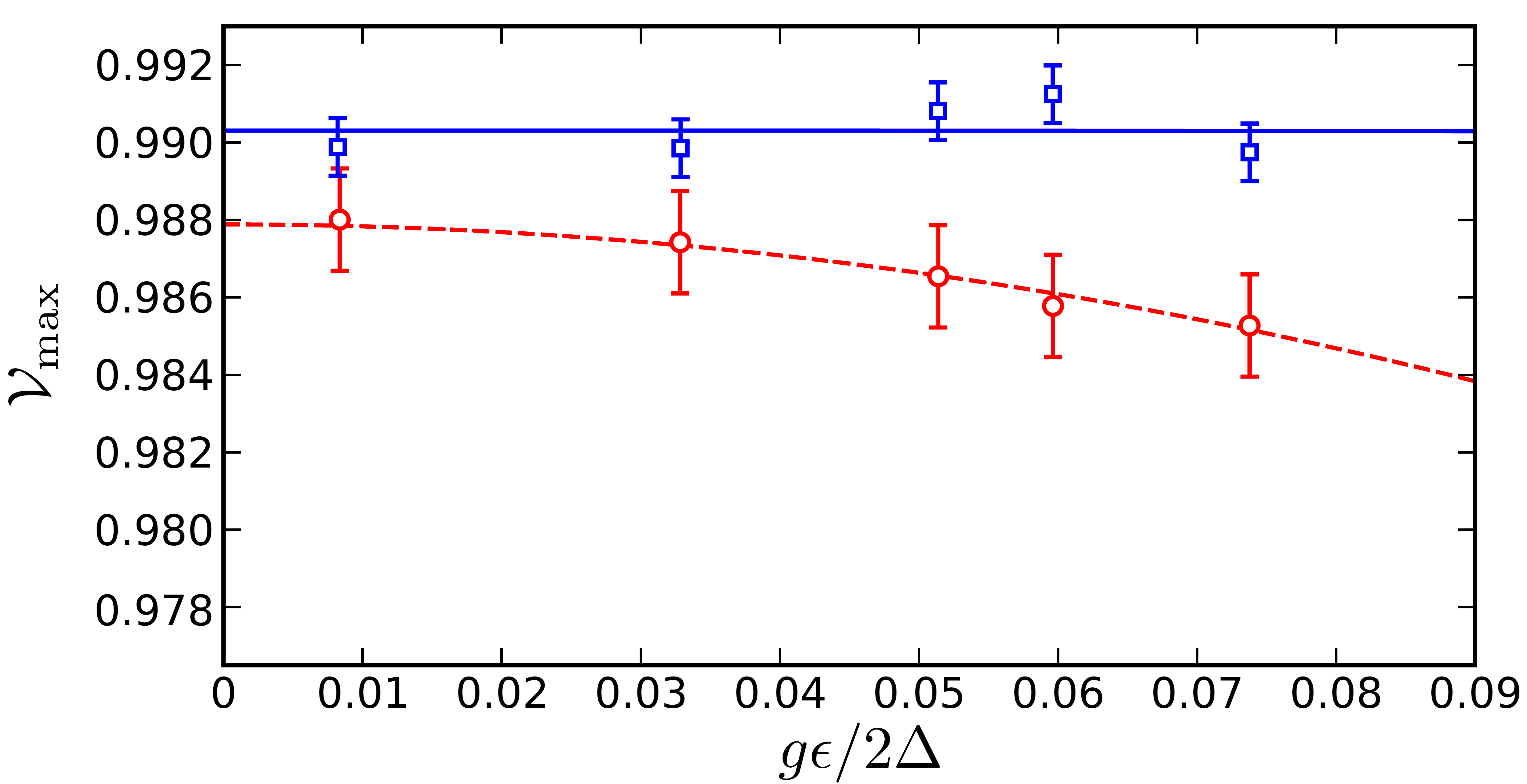}
\caption{
Maximally achievable visibility for different couplings $g\epsilon$ in the interference experiment between the undisturbed beam and the test beam (weak value case, blue boxes with blue solid line, and expectation value case, red circles with red dashed line) as a function of the coupling strength $g\epsilon/2\Delta$.
The lines represent the theoretical curves with fitted offsets.
Each setting was measured $10$ times to reduce the errors introduced by the optimization procedure.}
\label{exp_fig_1}
\end{figure}

Fig. \ref{exp_fig_1} presents our results for the observed visibility values for the two cases.
The action of the crystal leads to the intended (spatial) pointer shifts, resulting in the Bures angles $\frac{g\epsilon}{2\Delta}$ and $\frac{g^{2}\epsilon^{2}}{4\sqrt{2}\Delta^{2}}$ for the expectation value case and the weak value case, respectively.
The precision of the measurement is not sufficient to observe the tiny reduction of visibility in the weak value case when increasing $g\epsilon$ in the range available in the experiment given by the properties of the crystal.
However, the decrease of visibility for the expectation value case can clearly be resolved.

With real optical components, the maximal visibility is limited, in our experiment to about 99.0\% for $g\epsilon=0$, mainly due to coupling to higher order spatial modes.
These imperfections can be modeled by a fixed pointer shift with amplitude $\xi$ into a Hilbert space direction orthogonal to the initial pointer state.
Because of the high dimensionality of the Hilbert space of the pointer, it is unlikely that the imperfections are collinear to the intended shift and we assume them to be orthogonal.
Taking into account the influence of the imperfections, the dependences of the Bures angle given in (\ref{exp_angle_ex}) and (\ref{exp_angle_weak}) change to
\begin{eqnarray}\label{fitBuresExp}
D_{A}\left(\Phi_{{\rm e}},\rho_{{\rm ex}}\right) &= & \sqrt{\xi_1^2+\left(\frac{g\epsilon}{2\Delta}\right)^2}, \\
D_{A}\left(\Phi_{{\rm e}},\Phi_{{\rm w}}\right) &= & \sqrt{\xi_2^2+\left(\frac{g^{2}\epsilon^{2}}{4\sqrt{2}\Delta^{2}}\right)^2},\label{fitBuresWeak}
\end{eqnarray}
respectively.
Due to slight differences in the setup, the parameters $\xi_1$ and $\xi_2$, i.e., the offsets are not exactly equal.

\begin{figure}[h]\label{Exp2}
\centering
\includegraphics[width=0.48\textwidth]{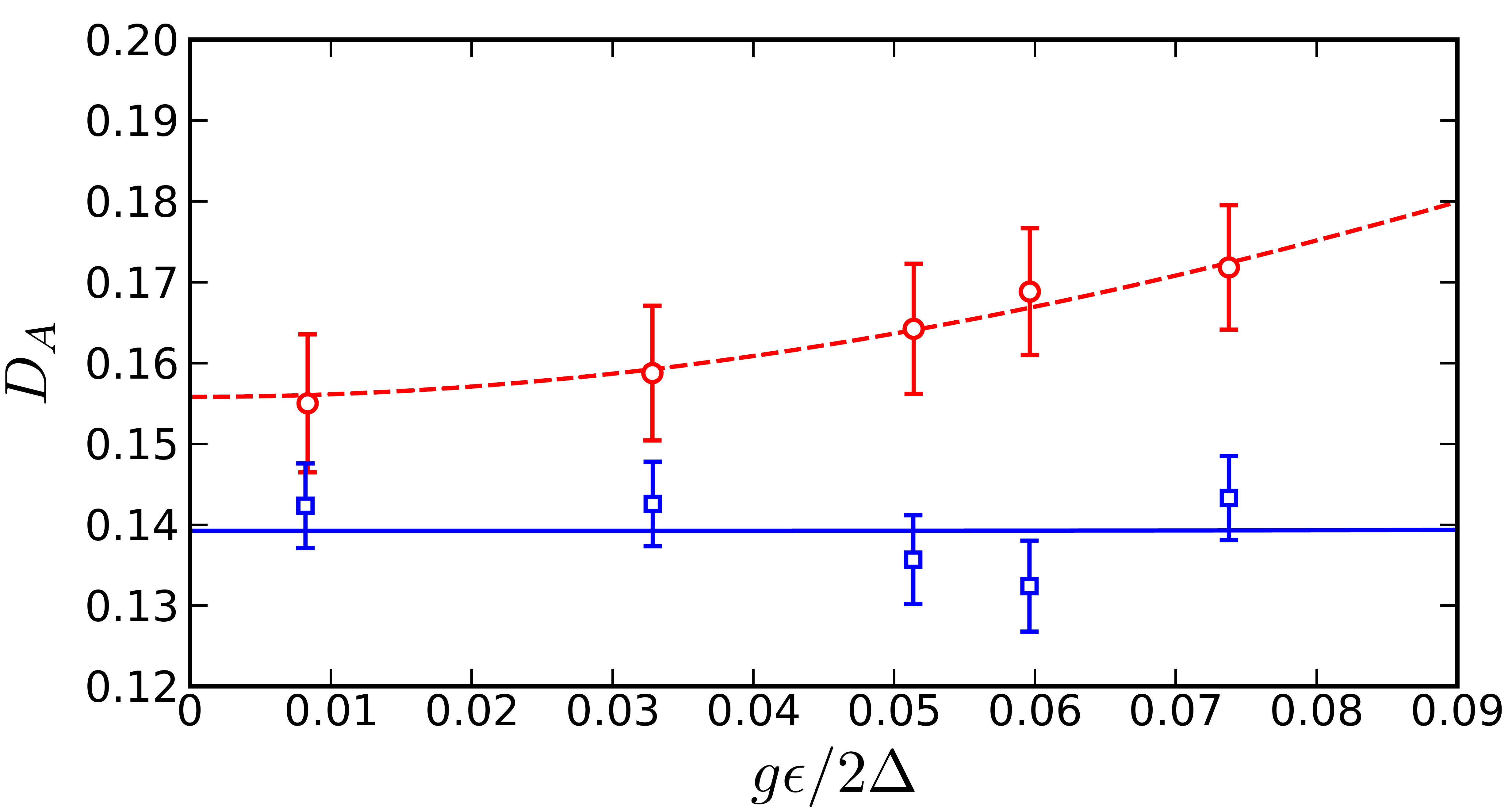}
\caption{Bures angles between the spatial state of the undisturbed beam (\ref{Phi0}) and the spatial states of the beam coupled to the weak value $A_w=0$ (blue boxes with blue solid fitted line) and the expectation value $\langle A \rangle=0$  (red circles with red dashed fitted line) as a function of the coupling strength $g\epsilon/2\Delta$.
The fitting according to equations (\ref{fitBuresWeak}) and (\ref{fitBuresExp}) includes offset parameters $\xi_1$ and $\xi_2$ appearing due to imperfections of the optical setup.}
\label{exp_fig_2}
\end{figure}

The Bures angles as functions of $g\epsilon$ for both cases are shown in Fig. \ref{exp_fig_2} together with the corresponding least squares fits according to (\ref{fitBuresExp}) and (\ref{fitBuresWeak}) with $\xi_1$ and $\xi_2$ being the only fit parameters.
We observe a good agreement with the theory (\ref{fitBuresExp}) and (\ref{fitBuresWeak}) in both cases.
While the Bures angle remains almost constant in the weak value case, it increases significantly for the expectation value case, which underpins our theoretical claim.

\section{Generalization of the result to arbitrary systems}\label{General}

A system with a weak value behaves in general in the same manner as a system described by an eigenvalue, not just for coupling to a Gaussian state through the Hamiltonian (\ref{VNInt}).
For the limit of an infinitesimally small time $\epsilon$, the  evolution of the system interacting with the pre- and post-selected system according to the Hamiltonian (\ref{VNInt}) is
\begin{equation}\label{shift}
\Phi (Q) =\Phi_0 (Q-g\epsilon A_w).
\end{equation}
This holds not just for Gaussian distributions, but for a wide class of wave functions with the Fourier transform reducing fast enough for large $P$.
Yet, it is not directly applicable for coupling to a pointer with a discrete spectrum (see Appendix $A$ for the case of a spin pointer).
What is generally true is that for a pre- and post-selected system the weak value (as an eigenvalue) should replace the corresponding operators in any interaction Hamiltonian, even when it is different from (\ref{VNInt}).
Note, however, that unlike eigenvalues, if in a non-linear Hamiltonian appears, say, $A^3$, it should not be replaced by $(A_w)^3$ but by the weak value of the respective operator, i.e., by the complex number $(A^3)_w$, which is in general different.
In Appendix $A$ we demonstrate the difference between coupling to systems described by weak values and systems described by expectation values for the case of a very different system than what was analyzed above, a spin$-\frac{1}{2}$ particle \cite{Molmer}.

The different scaling behavior demonstrated in the previous examples can be derived for arbitrary states and interactions.
Consider two systems, coupled by the Hamiltonian
\begin{equation}\label{VNInt3}
H_{\rm int}= gA B,
\end{equation}
where $A$ is a variable of the system, $B$ is a variable of the second system which we call pointer, and $g$ is a coupling constant.
The system is pre-selected in state $| \psi \rangle$ and post-selected in state $| \phi \rangle$.
The pointer system is prepared in state $|\Phi_0\rangle$.
Its state after the post-selection of the system is
\begin{equation}\label{w_state}
|\Phi_{\rm w}\rangle={\mathcal{N}} \langle\varphi | e^{-i g\epsilon A B} | \psi \rangle	|\Phi_0\rangle.
\end{equation}
If the system state $|\Psi\rangle$ is an eigenstate with eigenvalue $A=a$, numerically equal to the above weak value $A_w=a$, then the pointer ends up in the state
%If, instead, the second system is coupled to a system in an eigenstate with eigenvalue $A=a$, numerically equal to the weak value, $A_w=a$, then it ends up in the state
\begin{equation}\label{wv_eigenstate}
|\Phi_{\rm e}\rangle=e^{-i g\epsilon a B} |\Phi_0\rangle.
\end{equation}

Estimating the distance between these two states by expanding the exponents (including those in normalization factor $\mathcal{N}$) in powers of $\epsilon$ gives the Bures angle
\begin{equation}
\label{wv_scale}
D_A\left(\Phi_{\rm e},\Phi_{\rm w}\right) =
\left|\left(A^2\right)_w-a^2\right|\sqrt{\langle B^4 \rangle-\langle B^2 \rangle^2}~ \frac{g^2\epsilon^2}{2}+\mathcal{O}\left(\epsilon^3\right),
\end{equation}
where the variance $\sqrt{\langle B^4 \rangle-\langle B^2 \rangle^2}$ is computed at the time of pre-selection, i.e., for $|\Phi_0\rangle$.
Note that the first order of $\epsilon$ vanishes in all cases.

Now consider the first system to be pre-selected in the state $|\psi\rangle = \Sigma \alpha_k |a_k\rangle$ written in terms of eigenstates of $A$ without post-selection.
The expectation value $\langle A \rangle$ is assumed to be numerically equal to $a$.
We will compare the mixed state of the external system after the interaction
\begin{equation}\label{ex_state}
\rho_{\rm ex}=\Sigma \left|\alpha_k\right|^2 e^{-i g \epsilon a_k B} |\Phi_0\rangle\langle\Phi_0| e^{i g \epsilon a_k B},
\end{equation}
to the state $|\Phi_{\rm e}\rangle$ corresponding to coupling to the eigenvalue $a$ (actual or hypothetical).
Straightforward calculation based on
(\ref{overlap_mixture}) in which we expand in powers of $\epsilon$, yields the Bures angle
\begin{equation}\label{ex_scale}
D_A\left(\Phi_{\rm e},\rho_{\rm ex}\right)= \Delta A \Delta B g\epsilon + \mathcal{O}\left(\epsilon^2\right),
\end{equation}
where the uncertainties $\Delta A$ and $\Delta B$ are again those of the initial state.
In contrast to the weak value case the first order of $\epsilon$ does not drop out unless the system is pre-selected in an eigenstate of $A$ or if the pointer system is prepared in an eigenstate of $B$.

We have shown that the distance in the weak value case generally scales with a higher order than in the expectation value case.
Thus, in the weak interaction limit, the distance in the weak value case is infinitesimally small compared to the distance in the expectation value case.
In this limit we may say that the coupling specified by the weak value is identical to the coupling specified by an eigenvalue.
The coupling given by an expectation value is clearly different.
The application of (\ref{wv_scale}) and (\ref{ex_scale}) to several specific cases is considered in the Appendix $B$.

Care has to be taken for a complex weak value which cannot be an eigenvalue.
In this case the statement that in the interaction Hamiltonian the operator corresponding to a variable of a pre- and post-selected system should be replaced by the weak value remains correct.
Consequently, the effective Hamiltonian of systems coupled to a pre- and post-selected system is, in general, non-Hermitian \cite{nonhermitian}.
In Appendix $C$ we demonstrate the effect of a system with an imaginary weak value on other systems.
The scaling of the distance when the pre- and post-selected system characterized by a complex weak value is coupled to an arbitrary system is identical to that obtained for the real weak value case given by (\ref{wv_scale}).

\section{Generalization of weak values for mixed pre- and post-selected states}\label{stat_wv}

A natural way to generalize the expression for weak values (\ref{wv}) to density matrices is
\begin{equation}\label{wvm}
A_w\equiv\frac{\text{tr}\left(\rho_{\text{post}}A\rho_{\text{pre}}\right)}{\text{tr}\left(\rho_{\text{post}}\rho_{\text{pre}}\right)}.
\end{equation}
For pure pre- and post-selection states  the density matrices describing the system are $\rho_{\text{pre}}= |\psi\rangle\langle \psi |$, $\rho_{\text{post}}= |\varphi\rangle\langle \varphi|$, and the validity of this ansatz is immediately seen.
In the following, we will show that the expression (\ref{wvm}) for a weak value at time $t$ correctly shows the average effect of the system coupled to other systems at time $t$ through variable $A$ also for a case of a genuine \textit{mixed two-state vector}
 \begin{equation}\label{mixedTSV}
(\rho_{\text{post}}, \rho_{\text{pre}}).
\end{equation}

We do not expect that in a genuinely mixed case the weak value will be robust as an eigenvalue.
It can be shown that for external systems coupled to the system through variable $A$,
the deviation of the final states   from the final states in the case of numerically equal eigenvalue of $A$ is of order $\epsilon$. This is similar to the case of the expectation value.

In order to introduce the concept of a mixed two-state vector, we need to clarify how to pre- and post-select onto genuinely mixed states.
The concept of a mixed forward evolving  state is well understood.
The mixed  state $\rho_{\text{pre}}$ is obtained by preparation of a state entangled with an ancilla ($A_1$).
The state of the system remains mixed provided no measurement has been carried out on  $A_1$ after the preparation.
The future of the ancilla is unknown and this is what ensures that the state of the system is mixed.

In order to obtain a mixed backward evolving state we cannot just perform a post-selection measurement of a state entangled with another ancilla ($A_2$).
We need also to ensure that no measurement has been carried out on  $A_2$ {\it before} the post-selection measurement.
But since $A_2$ was created in a possibly known state, we have to  erase its past \cite{Vaidman2007}.

\begin{figure}[h]
  \centering
    \includegraphics[width=0.48\textwidth]{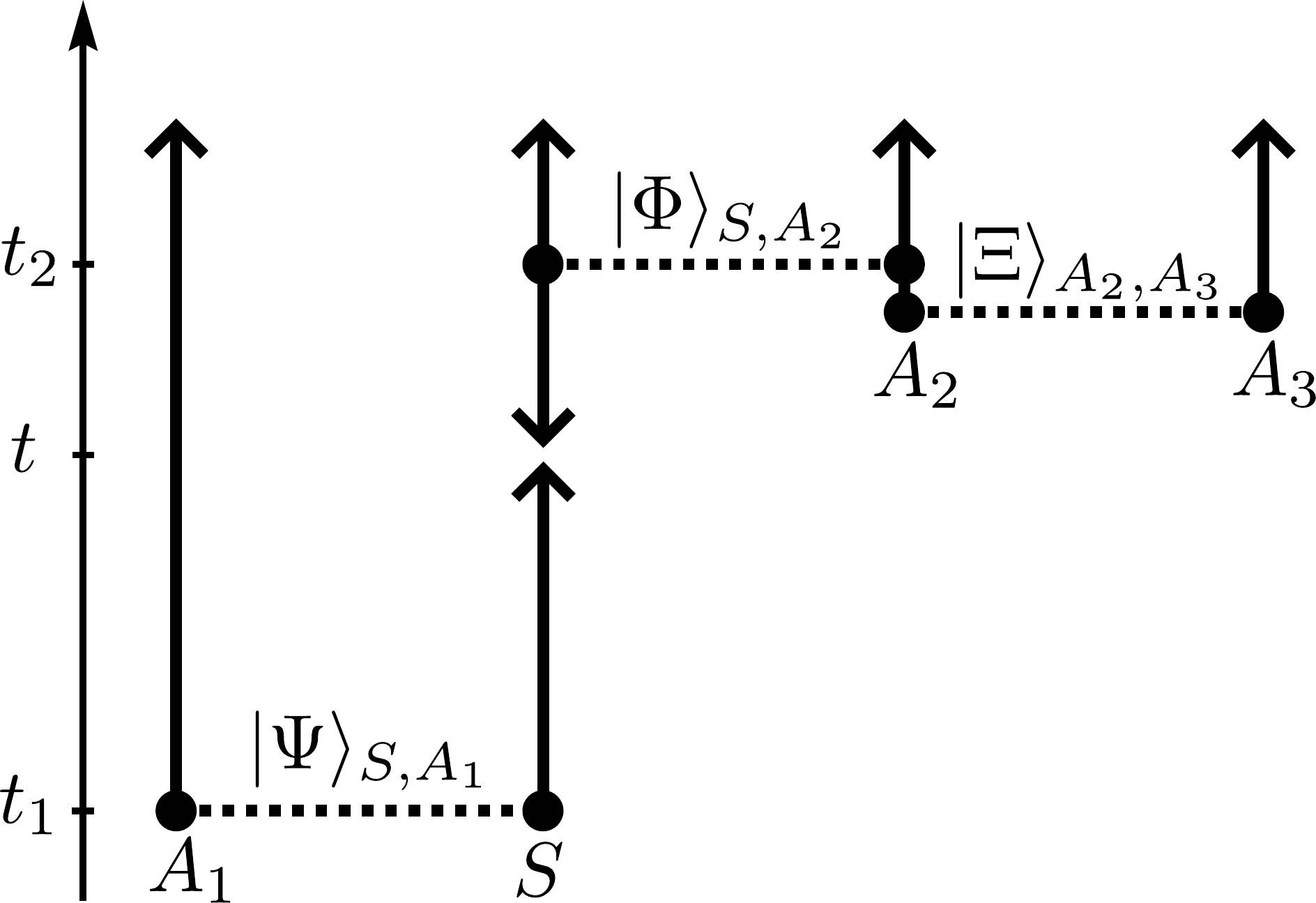}
  \caption{Preparation of a mixed two-state vector $(\rho_{\text{post}}, \rho_{\text{pre}})$. System $S$ and ancilla $A_1$ are prepared at $t_1$ in an entangled state $ |\Psi\rangle_{S,A_1}$ such that the reduced density matrix of $S$ is $\rho_{\text{pre}}$. The system is then found at $t_2$ in another entangled state $|\Phi\rangle_{S,A_2}$ with ancilla $A_2$ such that the reduced density matrix of $S$ is $\rho_{\text{post}}$. Just before $t_2$, ancilla $A_2$ was prepared in a maximally entangled state with yet another ancilla $A_3$. At intermediate times all ancilla systems are protected to avoid any interaction.}
  \label{mixed_diag}
\end{figure}

The scheme for the creation of a mixed two-state vector (\ref{mixedTSV}) at time $t$, $t_1<t<t_2$, is described in Fig.~\ref{mixed_diag}.
At time $t_1$ we prepare an entangled state with  $A_1$ and ensure that no measurement is performed on it after creation of the entangled state.
This provides a pre-selected mixed state.
Shortly before time $t_2$ we prepare a maximally entangled state of $A_2$ and ancilla $A_3$.
This erases the past of $A_2$ relative to time $t_2$ by connecting it to the future of  $A_3$ which can be ensured to be unknown.
(Another way to create a genuinely mixed backward evolving state might be realized by crossed in time nonlocal measurements, which were used in the first continuous variables teleportation scheme \cite{Vtele}, removing the need for the third ancilla.)

The pre-selected mixed state, described by density matrix
\begin{equation}
 \rho_{\rm pre}=  \sum_{k}  p_k|\psi_k\rangle \langle\psi_k|,
\end{equation}
 is created by preparing an entangled state with $A_1$:
 \begin{equation} \label{pre_mixed}
|\Psi\rangle_{S,A_1}=  \sum_{k}  \sqrt{p_k}|\psi_k\rangle|k\rangle_1.
\end{equation}

The post-selected mixed state, described by density matrix
\begin{equation}
\rho_{\rm post}=  \sum_{i}  \tilde{p}_i|\phi_i\rangle \langle\phi_i|,
\end{equation}
is created by first preparing a maximally entangled state between $A_2$  and $A_3$,
\begin{equation}
|\Xi\rangle_{A_2,A_3}= \frac{1}{\sqrt{N}} \sum_{i=1}^N  |i\rangle_2|i\rangle_3,
\end{equation}
and shortly afterwards, at time $t_2$, performing a post-selection measurement of the entangled state
\begin{equation}
 |\Phi\rangle_{S,A_2}=  \sum_{i}  \sqrt{\tilde{p}_i}|\phi_i\rangle|i\rangle_2.
\end{equation}

Straightforward application of the formula (\ref{wvm}) for a weak value of $A$ for this mixed two-state vector yields
\begin{equation}\label{gwv1}
A_w=\frac{\sum_{i,k}\tilde{p}_i p_k\langle \psi_k |\phi_i\rangle \langle \phi_i |A|\psi_k\rangle}{\sum_{i,k} \tilde{p}_i p_k|\langle \phi_i |\psi_k\rangle|^2}.
\end{equation}
In Appendix D we prove that this is the correct expression and by this show the validity of the expression for weak values (\ref{wvm}) for a mixed two-state vector.

Systems described by mixed states, with various ways of obtaining information from future measurements, have been discussed before \cite{Wiseman,SANDU0,SANDU,MolmerPred}.
What allowed us to derive, arguably, the simplest symmetrical expression  (\ref{wvm}), is our special procedure for introducing the backward evolving mixed state.

The weak value (\ref{wvm}) is not analogous to an eigenvalue in the sense described in the previous sections. What we have here is  a statistical weak value, as explained in Appendix D.
The pointer is in a mixed state similar to the expectation value case discussed before.

It is important to note that pre- and post-selected systems described by a {\it generalized two state vector} \cite{AV91}
\begin{equation}\label{tsv}
\sum_{k=1}^N \alpha_k \langle \phi_k |~|\psi_k\rangle,
\end{equation}
are not described by a genuine mixed two-state vector and the weak value for such a system is robust as for a system described by a pure two-state vector.

The generalized two-state vector arises when the system and ancillas are described together by a pure two-state vector.
It is postulated that between the pre- and post-selection the ancilla is isolated and there is coupling to the system only.
A simple construction for the generalized two-state vector (\ref{tsv}) is the following two-state vector describing the system and ancilla:
\begin{equation}\label{tsv+a}
\frac{1}{\sqrt{N}} \sum_{k=1}^N  \langle \phi_k |\langle k |~~~  \sum_{k=1}^N  \alpha_k|\psi_k\rangle|k\rangle,
\end{equation}
where $\{|k\rangle\}$ is an orthonormal basis of states of the ancilla.
Naively one can apply (\ref{wvm}) for the case of a generalized two-state vector, too.
Indeed, (\ref{tsv+a}) tells us that the system is prepared in a mixed state since it is entangled with an ancilla, and that it is also post-selected in some other mixed state.
But applying formula (\ref{wvm}) in this case is a mistake.
We cannot trace out the ancilla in preparation and in post-selection separately, because it is the same ancilla.
Indeed, from (\ref{tsv+a}) we obtain
\begin{equation}\label{rhopr,rhopo}
\rho_{\rm post}= \frac{1}{N}\sum_{k=1}^N  |\phi_k\rangle \langle \phi_k |,~~~~~\rho_{\rm pre}=  \sum_{k=1}^N  |\alpha_k|^2|\psi_k\rangle|\langle \psi_k|,
\end{equation}
Substituting in (\ref{wvm}) yields
\begin{equation}
A_w=\frac{\sum_{i,k} |\alpha_k|^2\langle \psi_k |\phi_i\rangle \langle \phi_i |A|\psi_k\rangle}{\sum_{i,k} |\alpha_k|^2|\langle \phi_i |\psi_k\rangle|^2},
\end{equation}
which is obviously different from the correct expression defined in \cite{AV91}
\begin{equation}\label{gwv}
A_w\equiv\frac{\sum \alpha_k\langle \phi_k |A|\psi_k\rangle}{\sum \alpha_k\langle \phi_k |\psi_k\rangle}.
\end{equation}

This weak value for the generalized two-state vector (\ref{tsv}) is equal to the standard weak value of the combined system with the two-state vector (\ref{tsv+a}), $A_w=(A\otimes I)_w$, as defined in (\ref{wv}).
For weak values defined for generalized two-state vectors, it should therefore also be true that for coupling to other systems during infinitesimal time, the system behaves as a system in an eigenstate.

\section{Weak values and weak measurements}

In this section we want to connect our results with the general literature on weak values which considers it as an outcome of weak measurements, and define a procedure for specifying the weak value for situations when the  post-selection measurement does not specify the backward evolving state of involved systems completely. In such cases the system is described neither by a pure two-state vector, nor by a generalized two-state vector, nor by a mixed two-state vector.
Such a situation occurs when the system is coupled  to an external system which is not post-selected.
Here, we will consider an example of a measurement procedure which would represent a weak measurement in the limit of weak coupling or small period of time between pre- and post-selection.
We, however, take a finite time $\tau$ and a finite strength of the interaction $g$.

We consider a spin variable (\ref{A=SZ}) pre- and post-selected in the same states as in (\ref{100}).
But while (\ref{100}) describes a two-state vector at a particular time, we take the forward evolving state from (\ref{100}) as the pre-selected state at time $t=0$ and the backward evolving state from (\ref{100}) as the post-selected state at some finite time $\tau$.
For simplicity, we consider a coupling of the system to a spin pointer,  prepared in the initial state $|\Phi_0\rangle=|{\uparrow}_x\rangle$, with interaction Hamiltonian (\ref{VNInt2}) as described in Appendix A.
Note that the final state of the spin pointer can only be read with a tomography analysis involving a large ensemble.

\begin{figure}[h]
  \centering
    \includegraphics[width=0.48\textwidth]{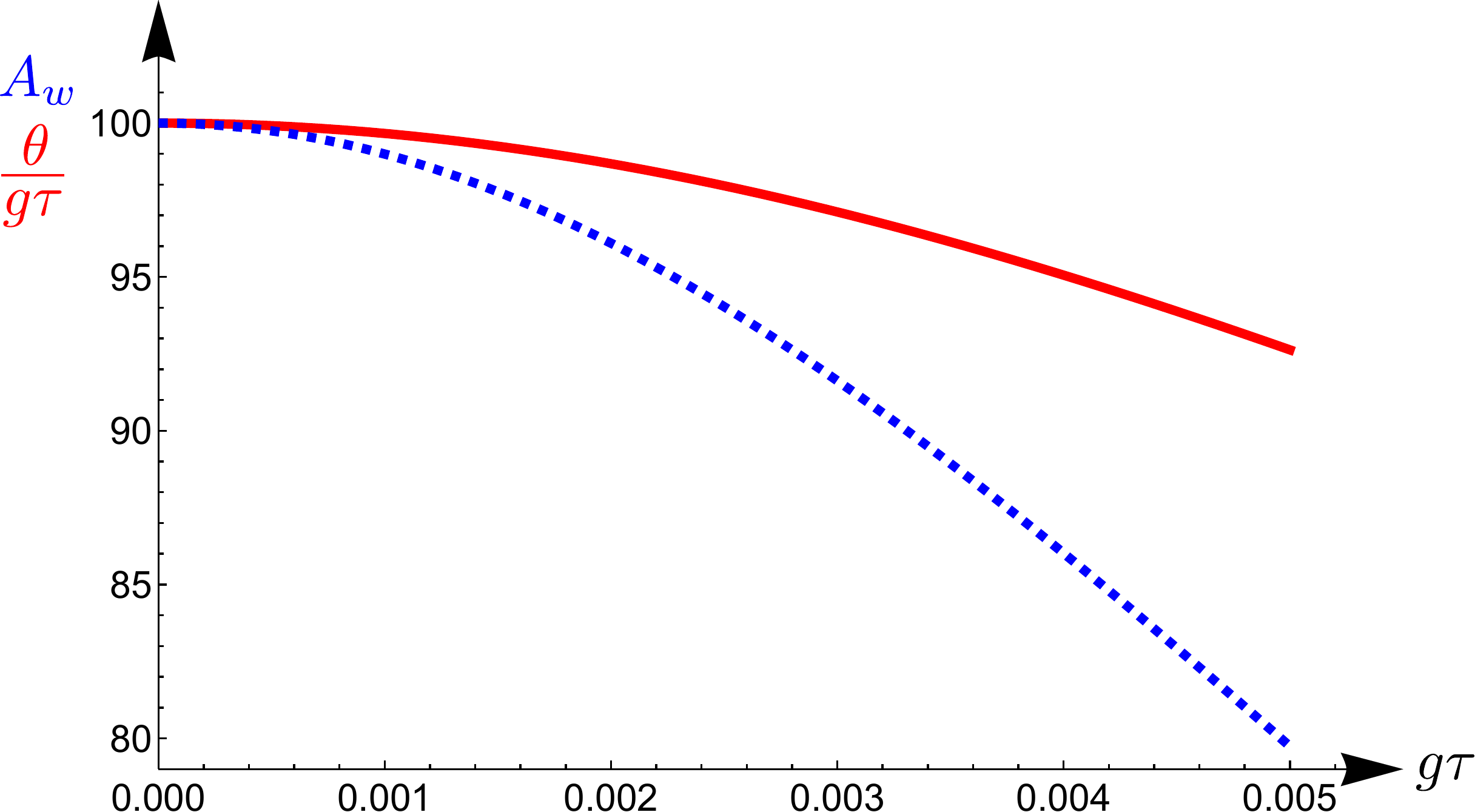}
  \caption{
  The pointer reading and the weak value as a function of the strength of the measurement interaction.
  The pointer reading, $\frac{\theta}{g\tau}$ (calibrated to correctly show eigenvalues of $A$) is given by the continuous red line.
  The weak value, $A_w$  is given by the blue dashed line. }
  \label{weak_angle_fig}
\end{figure}

If the interaction of this system with a spin pointer was very weak, we would expect the spin of the pointer to rotate by the angle $\theta =100 g\tau$ in proportion to the weak value $A_w=100$ corresponding to the two-state vector (\ref{100}).
For an arbitrary strength of the interaction the state of the spin and the pointer spin at time $\tau$ immediately before the post-selection is
\begin{equation} \label{spin-spin-t2}
\frac{1}{2} \left( |0\rangle |\uparrow\rangle + |0\rangle |\downarrow\rangle + e^{ig \tau}|-1\rangle |\uparrow\rangle + e^{-ig \tau}|-1\rangle |\downarrow\rangle \right).
\end{equation}
The state of the pointer spin at time $\tau$ immediately after the post-selection of the system is
\begin{equation} \label{ptr_fin}
|\Phi_{\tau}\rangle=\frac{(100e^{ig\tau}-101)|\uparrow\rangle+(100e^{-ig\tau}-101)|\downarrow\rangle}{\sqrt{40402-40400\cos(g\tau)}}.
\end{equation}
This is a pure state of the spin rotated around the $z$ axis.
Only in the limit of a weak measurement, i.e., $g\tau\rightarrow 0$, the angle $\theta$ of this rotation corresponds to $A=100$.
The continuous line on the plot in Fig.~\ref{weak_angle_fig} shows the dependence of the relative shift of the pointer variable $\frac{\theta}{g\tau}$ on the strength of the measurement interaction.

During the finite measurement interaction, the system couples to (and entangles with) an external system, which is not post-selected.
Therefore, the system itself is not described by a pure two-state vector, and thus, the basic definition (\ref{wv}) at an intermediate time $t$ cannot be applied.
The system is also not described by a mixed two-state vector if no special procedures as the one presented in the previous section are performed.
In order to find the weak value, we here present two approaches - a more general one and a special, but simpler procedure.
Our general approach uses the same ``trick'' as  in the proof of Appendix D. We rely on the fact that future operations cannot change any measurable property at present, and, in particular, the weak value at time $t$. Taking into account the assumption of  a vanishingly weak coupling at time $t$ to external systems, we can calculate the final state of the pointer. Then we introduce a verification measurement of this pointer state which we know  will effectively succeed  with certainty.
Therefore, the weak value, in case this measurement is successfully performed, has to be equal to the weak value in our experiment without the verification measurement.
With the measurement, the composite system, including the pointer, is described by a pure two-state vector which allows to calculate the generalized two-state vector of the system.
Finally, this gives the weak value at any intermediate time of any variable of the system.

The second method we present for calculating $A_w$  can be applied here because the variable $A$ commutes with the interaction Hamiltonian, and thus  $A_w$ at $t=\tau$, the time of the post-selection is $A_w$ during the whole period.
At time $t=\tau$, just before the post-selection, the composite system is described by mixed two-state vector (it is mixed only in the forward evolving state), so we can directly apply  (\ref{wvm}).
The entangled state of the system and the spin pointer just before time $\tau$ is (\ref{spin-spin-t2}), so the system is described by a forward evolving density matrix
\begin{equation} \label{den-mat-t2}
\rho_{\rm pre}=\begin{pmatrix}
0 & 0 & 0\\
0 & \frac{1}{2} & \frac{1}{2}\cos(g\tau) \\
0 & \frac{1}{2}\cos(g\tau) & \frac{1}{2} \\
\end{pmatrix}.
\end{equation}
The backward evolving state is given in (\ref{100}). The formula (\ref{wvm}) yields the weak value at $t=\tau$, and thus at all times:
\begin{equation} \label{wv-t}
A_w = \frac{201}{40402-40400\cos(g\tau)} -\frac{1}{2}.
\end{equation}
The dependence of the weak value on the strength of the measurement interaction is shown by the dotted line in Fig. \ref{weak_angle_fig}.

In the previous sections, we argued that the weak value of a physical variable at time $t$ characterizes an effective coupling to this variable at this moment.
At first glance, the discrepancy between the final reading of the spin pointer used in our measurement procedure and the weak value (which remains constant  during the interaction)  seem contradictory to the previous sections, since  we might expect that the pointer spin rotates as if the system has (hypothetical) eigenvalue equal $A_w$.
The reason for the discrepancy is the entanglement between the system and the pointer, so we have to consider the evolution of the whole composite system.
The weak value (\ref{wv-t}) remains relevant for  coupling of the system to external systems which are currently not entangled with our system.

\section{Conclusions}

We have analyzed the concept of the weak value.
Although it can be viewed as a statistical entity, i.e., the average reading of a measuring device, it has the deeper meaning of a property associated with a single quantum system similar to an eigenvalue when the quantum state of the system is a corresponding eigenstate.
We have shown theoretically and experimentally that the pre- and post-selected system coupled to external systems through a particular variable affects these systems at any infinitesimal period of time as if it were in an eigenstate corresponding to an eigenvalue numerically equal to the weak value.

In the experiment we observe different effects on the pointer for the cases of pre-selected photons (without post-selection) with expectation value $\langle A \rangle=0$ and pre- and post-selected photons with $A_w=0$.
This demonstrates that the nature of the weak value is different from the nature of the expectation value.
The measurements confirm that the weak value determines the interaction in the same way an eigenvalue does.

We have analyzed the concept of weak values for systems pre- and- post-selected in mixed states and derived a formula for weak values of variables of systems described by density matrices.
Evidently, due to the statistical nature of mixed states, the coupling to systems described by a mixed two-state vector is in general not equivalent to the coupling to an eigenvalue and the nature of weak values in this case has a genuine statistical element.

Our results deepen the understanding of the  behavior of quantum systems between measurements and can help to find new practical applications of quantum
experiments with pre- and post-selected systems.

\section*{Acknowledgments}

This work has been supported in part by the Israel Science Foundation Grant No. 1311/14,
the German-Israeli Foundation for Scientific Research and Development Grant No. I-1275-303.14,
and by the German excellence initiative NIM.
L.K. acknowledges support by the international PhD programme ExQM, J.D. by the international Max-Planck-Research school IMPRS-QST, and J.M. by a DAAD-fellowship.

\section*{Appendix A: Analysis of a measurement with a spin pointer}

The interaction Hamiltonian with a spin system is given by
\begin{equation}\label{VNInt2}
H_{\text{int}}=g A \sigma_z.
\end{equation}
We repeat the analysis of the cases of various pre- and post-selections of our system considered in section \ref{first_example}, but now coupled to the spin.
We take the initial state of the spin to be
\begin{equation}\label{phi0spin}
|\Phi_0\rangle=|{\uparrow}_x\rangle=\frac{1}{\sqrt{2}}(|{\uparrow}\rangle+|{\downarrow}\rangle ),
\end{equation}
which has maximal sensitivity to the coupling (\ref{VNInt2}).

When the system has the eigenvalue $A=1$, the coupling for time $\epsilon$ causes the rotation to the quantum state
\begin{equation}\label{feig}
|\Phi_{\rm e}\rangle=\frac{1}{\sqrt{2}}(e^{-i g\epsilon}|{\uparrow}\rangle+e^{i g\epsilon}|{\downarrow}\rangle ),
\end{equation}
such that
\begin{equation}\label{expectSpin}
D_A\left(\Phi_0,\Phi_{\rm e}\right)= g\epsilon.
\end{equation}

When the system is pre- and post-selected and described by the two-state vector (\ref{TSV}) corresponding to $A_w=1$,
the final state of the measuring device is
\begin{equation}
|\Phi_{\rm w}\rangle=\mathcal{N}[(2-e^{i g\epsilon})|{\uparrow}\rangle+(2-e^{-i g\epsilon})|{\downarrow}\rangle].
\end{equation}
This state is very close to the state given by (\ref{feig}) resulting in the Bures angle
\begin{equation}\label{DaSpin}
D_A\left(\Phi_{\rm e},\Phi_{\rm w}\right)=( g\epsilon)^3+\mathcal{O}(\epsilon^5).
\end{equation}

For the case of a system pre-selected in the state given in (\ref{psi3}) without post-selection, the density matrix representing the measuring device after the interaction is
\begin{equation}
\rho_{\rm ex}=\begin{pmatrix}
\frac{1}{2} & \frac{1}{4}(1+e^{-4i g\epsilon})\\
\frac{1}{4}(1+e^{4i g\epsilon}) & \frac{1}{2}
\end{pmatrix}.
\end{equation}
This mixed state is far from final state in case of coupling to the eigenvalue described by (\ref{feig}), which gives
\begin{equation} \label{spin_ex_dis}
D_A\left(\Phi_{\rm e},\rho_{\rm ex}\right)= g\epsilon+\mathcal{O}(\epsilon^3).
\end{equation}

Post-selection on the same state, which corresponds to the weak value $A_w=1$, leads to coupling which is again similar to the coupling to the eigenvalue $A=1$.
The distance between the states of the external system in these cases is proportional to $\epsilon^3$.
This is also true for the distance between the case $A_w=100$ obtained when the system is described by (\ref{100}) and the state of the system with eigenvalue $A=100$.
Thus, also in the case of coupling to a system very different from the Gaussian beam, described by a weak value and a numerically equal eigenvalue are the same (in the limit of weak interaction), while the coupling to a system described by a numerically equal expectation value is different.

 \section*{Appendix B: Consistency of the general results with examples considered in the paper}

In section IV we obtained general expressions for coupling to  systems described by weak values  (\ref{wv_scale}) and  expectation values (\ref{ex_scale}), respectively. Here we show that these results are consistent with analyses of the examples in previous sections.

In section \ref{first_example} the spin $A=S_z$ was measured by a continuous pointer according to (\ref{VNInt}) with $B=P$.
For the initial Gaussian pointer state we obtain
\begin{equation}
\langle P^2 \rangle = \frac{1}{4\Delta^2},~~\langle P^4 \rangle = \frac{3}{16\Delta^4}.
\end{equation}
For a spin described by the two-state vector (\ref{TSV}) it holds that
\begin{equation}\label{aA^2w}
\left(S_z\right)_w = a = 1,~~~ \left(S_z^2\right)_w = -1.
\end{equation}
Plugging these into (\ref{wv_scale}) yields (\ref{Phi2-Phi1}).

When the spin is described by the two-state vector (\ref{100}) we have
\begin{equation}
\left(S_z\right)_w = 100,~~~ \left(S_z^2\right)_w = -100,
\end{equation}
which allows one to obtain (\ref{DA_100}).

The formula (\ref{wv_scale}) works for coupling to the spin (\ref{VNInt2}) when $B=\sigma_z$ as well.
For the initial pointer spin state (\ref{phi0spin}), we obtain
\begin{equation}
\langle \sigma_{z}^2 \rangle = \langle \sigma_{z}^4 \rangle = 1.
\end{equation}
Together with (\ref{aA^2w}) it shows that (\ref{wv_scale}) is in agreement with (\ref{DaSpin}).

The general expressions are also confirmed by the examples in the expectation value case.
For example, considering the continuous pointer with $B=P$, coupled to the spin $A=S_z$ in the initial state (\ref{psi3}), yields the uncertainties
\begin{equation}
\Delta P = \frac{1}{2\Delta},~~ \Delta S_z = 1.
\end{equation}
Plugging these into (\ref{ex_scale}) results in an estimate of the distance consistent with (\ref{overlap_mixture}).

\section*{Appendix C: Weak values which are not real numbers}

For the coupling (\ref{VNInt}) to a continuous variable, the pointer wave function is ``shifted''  as in (\ref{shift}) even if the weak value is complex. Note, however, that the presence of the imaginary part in $A_w$ requires adding a normalization factor.
To demonstrate this behavior we will consider coupling to the photon polarization (\ref{polar}) with weak value $A_w=i$ obtained for initial state (\ref{pre}) and the post-selected state
\begin{equation}\label{post4}
\langle \phi |= \frac{1}{\sqrt 2}\left (\langle 1| +i \langle - 1| \right ).
\end{equation}
The pointer state after the post-selection is
\begin{equation}
\Phi_{\rm w}=\mathcal{N}
(e^{-\frac{(Q-g\epsilon)^{2}}{4\Delta^{2}}}-ie^{-\frac{(Q+g\epsilon)^{2}}{4\Delta^{2}}}),
\end{equation}
while the state  shifted by the imaginary ``eigenvalue" $i$ is
\begin{equation}
\Phi_{\rm e}=\mathcal{N}'e^{-\frac{(Q-ig\epsilon)^2}{4\Delta^{2}}}.
\end{equation}
Straightforward calculation shows that the distance between these  states  is small,
\begin{equation}\label{bures_complex_example}
D_A\left(\Phi_{\rm w},\Phi_{\rm e}\right)=\frac{ g^2 \epsilon^2}{2\sqrt{2}\Delta^2}+\mathcal{O}(\epsilon^4).
\end{equation}

If the coupling is to a spin variable (\ref{VNInt2}) instead of a continuous transverse degree of freedom, and we start again with the initial state of the measuring device $|\Phi_0\rangle=|{\uparrow}_x\rangle$, then the effective evolution of the spin is a rotation around the $y$ axis instead of the $z$ axis \cite{Hasegawa}.
It is approximately the same as the evolution under an effective non-Hermitian Hamiltonian in which the polarization operator is replaced by $i$.
The distance between the states in this case is
\begin{equation}
D_A\left(\Phi_{\rm w},\Phi_{\rm e}\right)=\frac{2 g^3 \epsilon^3}{3}+\mathcal{O}(\epsilon^5).
\end{equation}

\section*{Appendix D: Proof of the weak value formula for mixed states}\label{AppD}

We want to prove the expression for the weak value (\ref{wvm}) for a mixed two-state vector by deriving it from the basic definition (\ref{wv}).
However, our procedure, Fig.~\ref{mixed_diag}, does not provide a pure backward evolving state even for the composite system which includes our system and the three ancilla systems.
In order to resolve this issue we consider a verification measurement in the future, chosen such that it will have a definite result, thus providing the required backward evolving state.

\begin{figure}[h]
  \centering
    \includegraphics[width=0.48\textwidth]{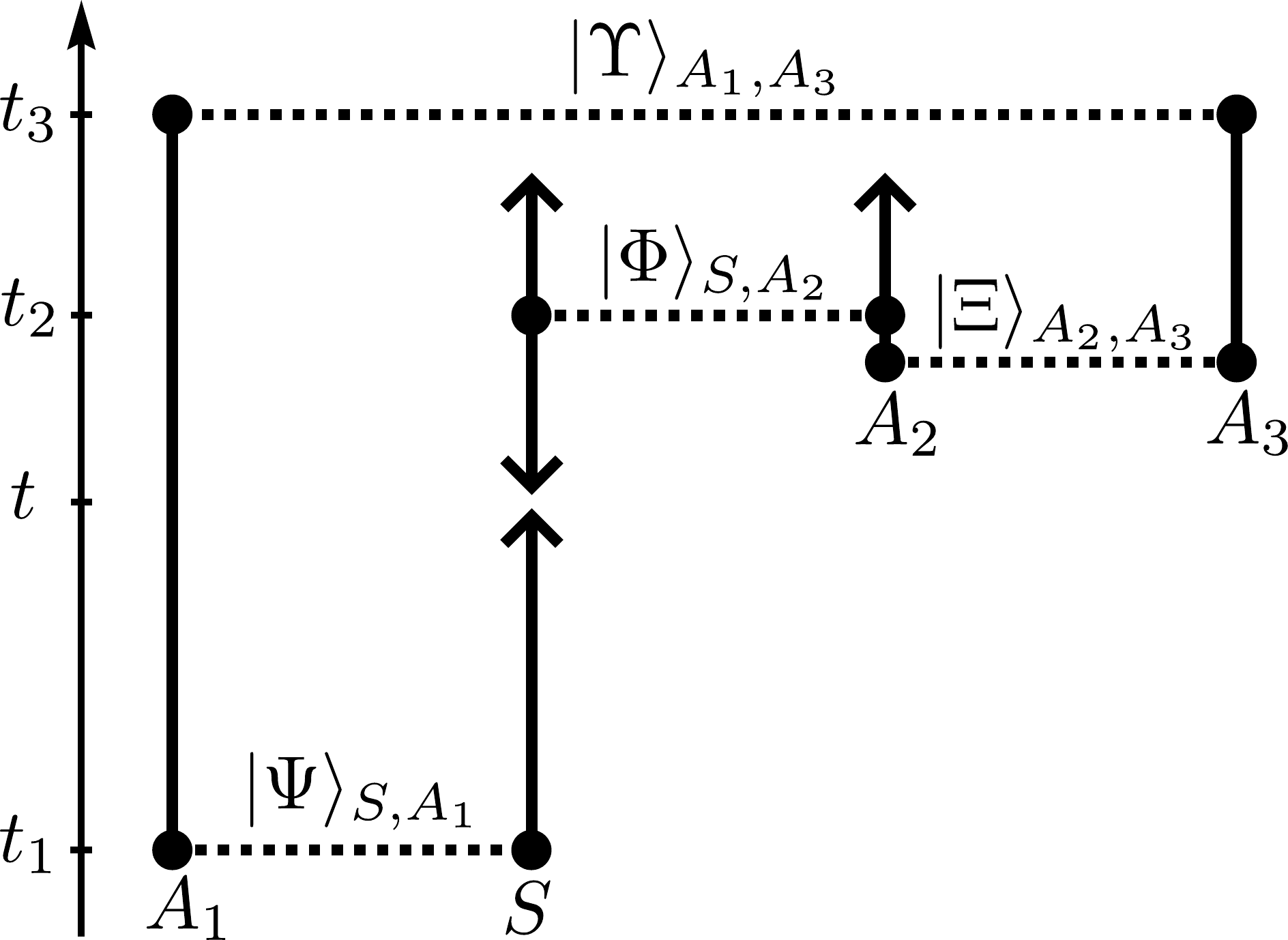}
  \caption{ In order to calculate the weak value, a (hypothetical) verification measurement is introduced at $t_3$ which provides  a pure two-state vector for the system and ancilla $A_1$ and thus a generalized two-state vector for the system.}
  \label{mixed_diag_appendix}
\end{figure}

The act of performing this verification measurement cannot change any measurable property at time $t$.
Having a large enough pre- and post-selected ensemble of systems (with their own ancilla systems) as described, the weak value of $A$ at time $t$ is a measurable property: the average shift of pointers weakly coupled at time $t$ to the systems in the pre- and post-selected ensemble.
In the limit where the weak coupling does not change the evolution, we can calculate the quantum state of the composite system after the whole process.
A verification measurement of this state at $t_3$ then will effectively succeed with certainty.
It will provide the pure backward evolving state with the same weak value as without the verification measurement, which is the weak value we wish to know.

Let us proceed with the proof.
At time $t_1$ we start with an entangled state of the system and ancilla $A_1$, $|\Psi\rangle_{S,A_1}$. Shortly before $t_2$ we prepare a maximally entangled state of ancillas $A_2$ and $A_3$,  $|\Xi\rangle_{A_2,A_3}$. Thus, before the post-selection measurement at time $t_2$, the total state of the system and the three ancilla systems is
\begin{equation}\label{preselection}
\left(\sum_{k}  \sqrt{p_k}|\psi_k\rangle|k\rangle_1\right) \left(\frac{1}{\sqrt{N}} \sum_{i=1}^N  |i\rangle_2|i\rangle_3\right).
\end{equation}
The post-selection measurement of the system and $A_2$ leaves ancilla systems $A_1$ and $A_3$ in the state
\begin{equation}
 |\Upsilon\rangle_{A_1,A_3} = \mathcal{N} \sum_{i,k}  \sqrt{{\tilde p}_i p_k}\langle \phi_i|\psi_k\rangle|k\rangle_1 |i\rangle_3.
\end{equation}
We consider a verification measurement at $t_3$ of state $|\Upsilon\rangle_{A_1,A_3}$, which  effectively succeeds with certainty.
Thus, the state of the system and the three ancilla systems evolving backward in time from  $t_2$ is
 \begin{equation}
\left(  \sum_{j}  \sqrt{{\tilde p}_j} \langle\phi_j|  \langle j |_2 \right) \left(\mathcal{N} \sum_{i,k}     \sqrt{{\tilde p}_i  p_k} \langle \psi_k|\phi_i\rangle \langle k|_1 \langle i|_3 \right).
\end{equation}
After evolving backwards through the Bell-type measurement of $A_2$ and $A_3$, the backward evolving state toward time $t$ is
 \begin{equation} \label{post_mixed}
\langle\Omega|_{S,A_1}=\mathcal{N} \sum_{i,k}  {\tilde p}_i \sqrt{p_k}\langle \psi_k|\phi_i\rangle \langle \phi_i| \langle k |_1.
\end{equation}
We have obtained a pure two-state vector of the system and $A_1$ at time $t$ with pre- and post-selected states given by (\ref{pre_mixed}) and (\ref{post_mixed}). This allows to apply (\ref{wv}) to find the weak value of $A$
\begin{eqnarray}  \label{aw_mixed_proof}
A_w &=& (A\otimes I_1)_w  = \frac{\langle\Omega| A\otimes I_1 |\Psi\rangle }{\langle\Omega|\Psi\rangle}  \\
&=&\frac{\sum_{i,j,k}  {\tilde p}_i \sqrt{p_j} \sqrt{p_k}\langle \psi_k|\phi_i\rangle \langle \phi_i| \langle k |_1 A\otimes I_1 |\psi_j\rangle|j\rangle_1}{\sum_{i,j,k}  {\tilde p}_i \sqrt{p_j} \sqrt{p_k}\langle \psi_k|\phi_i\rangle \langle \phi_i| \langle k |_1  |\psi_j\rangle|j\rangle_1}. \nonumber
\end{eqnarray}
This result reduces to the same expression as in (\ref{gwv1}), which provides the proof of our general formula (\ref{wvm}).

In our proof we have assumed that the verification measurement at $t_3$ effectively succeeds with certainty, which holds in the limit of vanishing interaction at time $t$.
This is enough for the proof since the weak value is defined at the limit, but the tiny probability for the failure of this measurement, which always exists, is crucial for the nature of the weak value of the genuinely mixed two-state vector.

The weak value (\ref{wvm}) is not analogous to an eigenvalue in the sense described in the previous sections.
Only the center of the affected pointer distribution shifts in the same manner as in the case of coupling to a system with a numerically equal eigenvalue.
In fact, it is the same situation as for the system described by an expectation value, where after the coupling the pointer is in the mixed state (\ref{ex_state}) .
The mixture with a tiny probability of an orthogonal pointer state (in case of the failure of the verification measurement) is equivalent to the mixture of almost identical states with comparable probabilities obtained via weak coupling to a system in a superposition of eigenstates.

  \end{document}